\documentstyle[12pt,lathuile,epsf,epsfig]{article}

\newcommand{\VEV}[1]{\left\langle{#1}\right\rangle}

\newcommand{\ie}{{\it i.e.}}
\newcommand{\eg}{{\it e.g.}}
\newcommand{\ms}{$\overline{\mbox{MS}}$}

\newcommand{\amst}{\mbox{${\widetilde{\alpha}_{\overline{\mbox{\tiny MS}}}}$}}
\newcommand{\av}{\mbox{$\alpha_{V}$}}
\newcommand{\GeV}{{\rm GeV}}
\begin{document}
\begin{flushright}
{\small
SLAC--PUB--8176\\
June 1999\\}
\end{flushright}

\begin{center}
{\bf\large   
CONFORMAL  SYMMETRY AS A TEMPLATE: \\
COMMENSURATE SCALE RELATIONS AND \\[1ex]
PHYSICAL RENORMALIZATION SCHEMES\footnote{\baselineskip=13pt
Work partially supported by the Department
of Energy, contract DE--AC03--76SF00515.}}

\medskip

Stanley J. Brodsky and Johan Rathsman        \\
{\em Stanford Linear Accelerator Center \\
 Stanford, California 94309} \\
\medskip

\end{center}

\vfill

\begin{center}
{\bf\large   
Abstract }
\end{center}

Commensurate scale relations are
perturbative QCD predictions which relate observable to observable at
fixed relative scale, such as the ``generalized Crewther relation",
which connects the Bjorken and Gross-Llewellyn Smith deep inelastic
scattering sum rules to measurements of the $e^+ e^-$ annihilation cross
section.  We show how conformal symmetry provides a template for such
QCD predictions, providing relations between observables which are
present even in theories which are not scale invariant.  All
non-conformal effects are absorbed by fixing the ratio of the respective
momentum transfer and energy scales.  In the case of fixed-point
theories, commensurate scale relations relate both the ratio of
couplings and the ratio of scales as the fixed point is approached.  In
the case of the $\alpha_V$ scheme defined from heavy quark interactions,
virtual corrections due to fermion pairs are analytically
incorporated into the Gell-Mann Low function, thus avoiding the problem
of explicitly computing and resumming quark mass corrections related to
the running of the coupling.  Applications to the decay width of the $Z$
boson, the BFKL pomeron, and virtual photon scattering are discussed.

\vfill

\begin{center} 
{\it Invited talk presented at Les Rencontres de Physique
de la Vall\'ee d'Aoste} \\
{\it La Thuile, Aosta Valley, Italy, February 28--March 6, 1999 }\\
\end{center}

\vfill

\newpage

\section{Introduction}

Testing quantum chromodynamics to high precision is not easy.  Even in
processes involving high momentum transfer, perturbative QCD predictions
are complicated by questions of the convergence of the series,
particularly by the presence of ``renormalon" terms which grow as $n!$,
reflecting the uncertainty in the analytic form of the QCD coupling at
low scales.  Virtually all QCD processes are complicated by the presence
of dynamical higher twist effects, including power-law suppressed
contributions due to multi-parton correlations, intrinsic transverse
momentum, and finite quark masses.  Many of these effects are inherently
nonperturbative in nature and require knowledge of hadron wavefunction
themselves.  The problem of interpreting perturbative QCD predictions is
further compounded by theoretical ambiguities due to the apparent
freedom in the choice of renormalization schemes, renormalization
scales, and factorization procedures.

A central principle of renormalization theory is that predictions which
relate physical observables to each other cannot depend on theoretical
conventions.  For example, one can use any renormalization scheme, such
as the modified minimal subtraction scheme,
and any choice of renormalization scale $\mu$ to compute perturbative
series relating observables $A$ and $B$.  However, all traces of the
choices of the renormalization scheme and scale must disappear when one
algebraically eliminates the $\alpha_{\overline{\mbox{\tiny MS}}}(\mu)$
and directly relate $A$ to $B$.  This is the principle underlying
``commensurate scale relations" (CSR),\cite{CSR} which are general
leading-twist QCD predictions relating physical observables to each
other.  For example, the ``generalized Crewther relation", which is
discussed in more detail below, provides a scheme-independent relation
between the QCD corrections to the Bjorken (or Gross Llewellyn-Smith)
sum rule for deep inelastic lepton-nucleon scattering, at a given
momentum transfer $Q$, to the radiative corrections to the annihilation
cross section $\sigma_{e^+ e^- \to \rm hadrons}(s)$, at a corresponding
``commensurate" energy scale $\sqrt s$.  \cite{CSR,BGKL} The specific
relation between the physical scales $Q$ and $\sqrt s$ reflects the fact
that the radiative corrections to each process have distinct quark mass
thresholds.

Any perturbatively calculable physical quantity can be used to define an
effective charge\cite{Grunberg,DharGupta,GuptaShirkovTarasov} by incorporating
the entire radiative correction into its definition.
For example, the $e^+ e^-\to  \gamma^* \to {\rm
hadrons}$ to muon pair cross section ratio can be written
$(C_F = (N^2_C-1)/2N_C)$
\begin{equation}
R_{e^+ e^-}(s) \equiv R^0_{e^+ e^-}(s) \left[ 1 +
\frac{3}{4}C_F{\alpha_R(s)\over \pi} \right] ,
\end{equation}
where $R^0_{e^+ e^-}$ is the prediction at Born level.
Similarly, one can define the entire radiative correction to the Bjorken
sum rule as the effective charge $\alpha_{g_1}(Q^2)$ where
$Q$ is the corresponding momentum transfer:
\begin{equation}
\int_0^1 d x \left[ g_1^{ep}(x,Q^2) - g_1^{en}(x,Q^2) \right]
   \equiv {1\over 6} \left|g_A \over g_V \right|
   C_{\rm Bj}(Q^2)
 = {1\over 6} \left|g_A
\over g_V \right| \left[ 1- {3\over 4} C_F{\alpha_{g_1}(Q^2) \over
   \pi} \right]  .
\end{equation}

By convention, each effective charge is normalized to $\alpha_s$ in the
weak coupling limit.  One can define effective charges for virtually any
quantity calculable in perturbative QCD; \eg~moments of structure
functions, ratios of form factors, jet observables, and the effective
potential between massive quarks.  In the case of decay constants of the
$Z$ or the $\tau$, the mass of the decaying system serves as the
physical scale in the effective charge.  In the case of multi-scale
observables, such as the two-jet fraction in $e^+ e^-$ annihilation, the
multiple arguments of the effective coupling $\alpha_{2 jet}(s,y)$
correspond to the overall available energy $s$ and variables such as $y
= \rm {max}_{ij} (p_i+p_j)^2/s$ representing the maximum jet mass
fraction.

Commensurate scale relations take the general form
\begin{equation}
\alpha_A(Q_A) = C_{AB}[\alpha_B(\Lambda_{BA}Q_A)]
=
\alpha_B(\Lambda_{BA}Q_A) +
\sum_{n=1}^{\infty} c_n^{AB}\frac{\alpha_B^{n+1}}{\pi^n}(\Lambda_{BA}Q_A)
\ .
\end{equation}
The function $C_{AB}(\alpha_B)$ relates the observables $A$ and $B$ in
the conformal limit; \ie, $C_{AB}$ gives the functional dependence
between the effective charges which would be obtained if the theory had
zero $\beta$ function.  The conformal coefficients $c_n^{AB}$
can be distinguished
from the terms associated with the $\beta$ function at each order in
perturbation theory from their color and flavor dependence, or by an
expansion about a fixed point.

The ratio of commensurate scales $\Lambda_{BA}$ is determined by the
requirement that all terms involving the $\beta$ function are
incorporated into the arguments of the running couplings, as in the
original BLM procedure.\cite{BLM} Physically, the ratio of scales
corresponds to the fact that the physical observables have different
quark threshold and distinct sensitivities to fermion loops.  More
generally, the differing scales are in effect relations between mean
values of the physical scales which appear in loop integrations.
Commensurate scale relations are transitive; \ie, given the relation
between effective charges for observables $A$ and $C$ and $C$ and $B$,
the resulting between $A$ and $B$ is independent of $C$.  In particular,
transitivity implies $\Lambda_{AB} = \Lambda_{AC} \times \Lambda_{CB}$.
The shift in scales which gives conformal coefficients in effect
pre-sums the large and strongly divergent terms in the PQCD series which
grow as $n!  (\beta_0 \alpha_s)^n$, \ie, the infrared renormalons
associated with coupling-constant
renormalization.\cite{tHooft,Mueller,LuOneDim,BenekeBraun}

One can consider QCD predictions as functions of analytic variables of
the number of colors $N_C$ and flavors $N_F$.  For example, one can show
at all orders of perturbation theory that PQCD predictions reduce to
those of an Abelian theory at $N_C \to 0$ with $C_F \alpha_s$ and
$N_F/T_FC_F$ held fixed, where $C_F=4/3$ and $T_F=1/2$ for QCD.  In
particular, CSRs obey the ``Abelian correspondence principle" in that
they give the correct Abelian relations at $N_C \to 0.$
\cite{BLM,BrodskyHuet,LepageMackenzie,Neubert,BallBenekeBraun}

Similarly, commensurate scale relations obey the ``conformal
correspondence principle":  the CSRs reduce to correct conformal
relations when $N_C$ and $N_F$ are tuned to produce zero $\beta$
function.  Thus conformal symmetry provides a {\it template} for QCD
predictions, providing relations between observables which are present
even in theories which are not scale invariant.  All effects of the
nonzero beta function are encoded in the appropriate choice of relative
scales $\Lambda_{AB} = Q_A/Q_B$.

The scale $Q$ which enters a given effective charge corresponds to a
physical momentum scale.  The total logarithmic derivative of each
effective charge $\alpha_A(Q)$ with respect to its
physical scale is given by the Gell Mann-Low equation:  
\begin{equation} {d\alpha_A(Q,m) \over d \log Q} = \Psi_A
(\alpha_A(Q,m), Q/m) , \end{equation} 
where the functional dependence
of $\Psi_A$ is specific to its own effective charge.  Here $m$ refers to
the quark's pole mass.  The pole mass is universal in that it does not
depend on the choice of effective charge.  The Gell Mann-Low relation is
reflexive in that $\Psi_A$ depends on only on the coupling $\alpha_A$ at
the same scale.  It should be emphasized that the Gell Mann-Low equation
deals with physical quantities and is independent of the renormalization
procedure and choice of renormalization scale.  A central feature of
quantum chromodynamics is asymptotic freedom; \ie, the monotonic
decrease of the QCD coupling $\alpha_A(Q)$ at large spacelike scales.
The empirical test of asymptotic freedom is the verification of the
negative sign of the Gell Mann-Low function at large momentum transfer,
which must be true for any effective charge.

In perturbation theory, 
\begin{equation} \Psi_A = - \psi_A^{\{0\}}
{\alpha_A^2\over \pi} - \psi_A^{\{1\}}{\alpha_A^3\over \pi^2} -
\psi_A^{\{2\}} {\alpha_A^4\over \pi^3} + \cdots \; . \end{equation} 
At
large scales $Q^2 \gg m^2$, the first two terms are universal and
identical to the first two terms of the $\beta$ function
$\psi_A^{\{0\}}=\beta_0 = {11 N_C \over 6} - {2 \over 3} T_F N_F,
\psi_A^{\{1\}}=\beta_1={17 \over 12} C_A^2 - {5\over 6} C_A T_F N_F - {1
\over 2} C_F T_F N_F, $ whereas $\psi_A^{\{n\}}$ for $n \ge 2$ are
process dependent.  The quark mass dependence of the $\Psi$ function is
analytic, and in the case of the $\alpha_V$ scheme is known to two
loops.\cite{bmr} Collecting all the mass effects into a mass dependent
function $N_F$ gives the effective number of flavors in the V-scheme;
$\psi_{V}^{\{0\}}\left( Q/m \right) =
\frac{11}{2}-\frac{1}{3}N_{F,V}^{\{0\}}\left( Q/m \right)$ and
$\psi_{V}^{\{1\}} \left( Q/m \right) = \frac{51}{4}
-\frac{19}{12}N_{F,V}^{\{1\}}\left( Q/m \right) $, with the subscript
$V$ indicating the scheme dependence.

The commensurate scale relation between $\alpha_A$ and $\alpha_B$
implies an elegant relation between their conformal dependence $C_{AB}$
and their respective Gell Mann Low functions:  
\begin{equation} \Psi_B
= {d C_{BA}\over d \alpha_A} \times {d \Lambda_{AB}\over d \log Q_B}
\times \Psi_A . \end{equation} 
Thus given the result for
$N_{F,V}(m/Q)$ in the $\alpha_V$ scheme one can use the CSR to derive
$N_{F,A}(m/Q)$ for any other effective charge, at least to two loops.
The above relation also shows that if one effective charge has a fixed
point $\Psi_A[\alpha_A(Q^{FP}_A)] = 0$, then all effective charges $B$
have a corresponding fixed point $\Psi_B[\alpha_B(Q^{FP}_B)] = 0$ at the
corresponding commensurate scale and value of effective charge.

In quantum electrodynamics, the running coupling $\alpha_{QED}(Q^2)$,
defined from the Coulomb scattering of two infinitely heavy test charges
at the momentum transfer $t = -Q^2$, is taken as the standard
observable.  Is there a preferred effective charge which one should use
to characterize the coupling strength in QCD?  In the case of QCD, the
heavy-quark potential $V(Q^2)$ is customarily defined via a Wilson loop from the
interaction energy of infinitely heavy quark and antiquark at momentum
transfer $t = -Q^2.$ The relation $V(Q^2) = - 4 \pi C_F
\alpha_V(Q^2)/Q^2$ then defines the effective charge $\alpha_V(Q).$ As
in the corresponding case of Abelian QED, the scale $Q$ of the coupling
$\alpha_V(Q)$ is identified with the exchanged momentum.  Thus there is
never any ambiguity in the interpretation of the scale.  All virtual corrections 
due to fermion pairs are incorporated in
$\alpha_V$ through loop diagrams which depend on
the physical mass thresholds.  Other observables could be used to define
the standard QCD coupling, such as the effective charge defined from
heavy quark radiation.\cite{Uraltsev}

Commensurate scale relations between $\alpha_V$ and the QCD radiative
corrections to other observables have no scale or scheme ambiguity, even
in multiple-scale problems such as multi-jet production.  As is the case
in QED, the momentum scale which appears as the argument of $\alpha_V$
reflect the mean virtuality of the exchanged gluons.  Furthermore, one
can write a commensurate scale relation between $\alpha_V$ and an
analytic extension of the $\alpha_{\overline{\mbox{\tiny MS}}}$
coupling, thus transferring most of the unambiguous scale-fixing and
analytic properties of the physical $\alpha_V$ scheme to the
$\overline{\mbox{MS}}$ coupling.

Some examples of CSR's at NNLO are given by:
\begin{equation}
\alpha_R(\sqrt s) =
\alpha_{g_1}(0.5 \sqrt s) -
{\alpha^2_{g_1}(0.5 \sqrt s)\over \pi} +
{\alpha^3_{g_1}(0.5 \sqrt s)\over \pi^2}
\end{equation}
\begin{equation}
\alpha_R(\sqrt s) =
\alpha_V(1.8 \sqrt s) + 2.08
{\alpha^2_V(1.8 \sqrt s)\over \pi} - 7.16
{\alpha^3_V(1.8 \sqrt s)\over \pi^2}
\end{equation}
\begin{equation}
\alpha_{\tau}(m_\tau) =
\alpha_V(0.8 m_\tau) + 2.08
{\alpha^2_V(0.8 m_\tau)\over \pi} - 7.16
{\alpha^3_V(0.8 m_\tau)\over \pi^2}
\end{equation}
\begin{equation}
\alpha_{g_1}(Q) =
\alpha_V(0.8 Q) + 1.08
{\alpha^2_V(0.8 Q)\over \pi} - 10.3
{\alpha^3_V(0.8 Q)\over \pi^2}
\end{equation}
For numerical purposes in each case $N_F=5$ and $\alpha_V=
0.1$ have been used to compute the NLO correction to the CSR scale.

Commensurate scale relations thus provide fundamental and precise
scheme-independent tests of QCD, predicting how observables track not
only in relative normalization, but also in their commensurate scale
dependence.

\section{The Generalized Crewther Relation}

The generalized Crewther relation\cite{BGKL} can be derived by
calculating the QCD radiative corrections to the deep inelastic sum
rules and $R_{e^+ e^-}$ in a convenient renormalization scheme such as
the modified minimal subtraction scheme $\overline{\rm MS}$.  One then
algebraically eliminates $\alpha_{\overline{\mbox{\tiny MS}}}(\mu)$.
Finally, BLM scale-setting\cite{BLM} is used to eliminate the
$\beta$-function dependence of the coefficients.  The form of the
resulting relation between the observables thus matches the result which
would have been obtained had QCD been a conformal theory with zero
$\beta$ function.  The final result relating the observables is
independent of the choice of intermediate $\overline{\rm MS}$
renormalization scheme.

More specifically, consider the Adler function\cite{Adler} for the $e^+
e^-$ annihilation cross section
\begin{equation} D(Q^2)=-12\pi^2 Q^2{d\over dQ^2}\Pi(Q^2),~
\Pi(Q^2) =-{Q^2\over 12\pi^2}\int_{4m_{\pi}^2}^{\infty}{R_{e^+ e^-}(s)ds\over
s(s+Q^2)}.
\end{equation}
The entire radiative correction to this function is defined as
the effective charge
$\alpha_D(Q^2)$:
\begin{eqnarray}
    D \left( Q^2\right)
&\equiv&
    3 \sum_f Q_f^2 \left[ 1+ {3\over 4} C_F{\alpha_D(Q^2) \over \pi}
                   \right]
    +( \sum_f Q_f)^2C_{\rm L}(Q^2) \nonumber \\
&\equiv& 3 \sum_f Q_f^2 C_D(Q^2)+( \sum_f Q_f)^2C_{\rm L}(Q^2),
\nonumber
\end{eqnarray}
The coefficient $C_{\rm L}(Q^2)$ appears at the third order in
perturbation theory and is related to the ``light-by-light scattering type"
diagrams.

It is straightforward to algebraically relate $\alpha_{g_1}(Q^2)$ to
$\alpha_D(Q^2)$ using the known expressions to three loops in the
$\overline{\rm MS}$ scheme.  If one chooses the renormalization scale to
resum all the  corrections due to the running of the coupling  into
$\alpha_D(Q^2)$, then the final result turns out to be remarkably
simple\cite{BGKL} $(\widehat\alpha = 3/4\, C_F\ \alpha/\pi):$
\begin{equation}
\widehat{\alpha}_{g_1}(Q)=\widehat{\alpha}_D( Q^*)-
\widehat{\alpha}_D^2( Q^*)+\widehat{\alpha}_D^3( Q^*) + \cdots,
\end{equation}
where
\begin{eqnarray}
\ln { {Q}^{*} \over Q} &=& {7\over
4}-2\zeta(3)+ \frac{\alpha_D ( Q^*)}{\pi} \Biggl[ \left( {11\over
48}+{14\over 3} \zeta(3)-4{\zeta^2(3)} \right) \beta_0\cr && +{13\over
36}C_{\rm A} -{1\over 3}C_{\rm A}\zeta(3) -{145\over 144} C_{\rm F}
-{23\over 3}C_{\rm F}\zeta(3) +10C_{\rm F}\zeta(5) \Biggr] \\ &=&
-0.654+\frac{\alpha_D ( Q^*)}{\pi}(0.059\beta_0+0.0767) \ .
\label{EqLogScaleRatio}
\end{eqnarray}
This relation shows how the coefficient functions for these two
different processes are related to each other at their respective
commensurate scales.  We emphasize that the $\overline{\mbox{MS}}$
renormalization scheme is used only for calculational convenience; it
serves simply as an intermediary between observables.  The
renormalization group ensures that the forms of the CSR relations in
perturbative QCD are independent of the choice of an intermediate
renormalization scheme.

The Crewther relation was originally derived assuming that the theory is
conformally invariant; \ie, for zero $\beta$ function.  In the physical
case, where the QCD coupling runs, all non-conformal effects are
resummed into the energy and momentum transfer scales of the effective
couplings $\alpha_R$ and $\alpha_{g1}$.  The general relation between
these two effective charges for non-conformal theory thus takes the form
of a geometric series
\begin{equation}
        1- \widehat \alpha_{g_1} =
\left[ 1+ \widehat \alpha_D( Q^*)\right]^{-1} \ .
\end{equation}
We have dropped the small light-by-light scattering contributions.  This
is again a special advantage of relating observable to observable.  The
coefficients are independent of color and are the same in Abelian,
non-Abelian, and conformal gauge theory.  The non-Abelian structure of
the theory is reflected in the expression for the scale ${Q}^{*}$.

Is experiment consistent with the generalized Crewther relation?  Fits
\cite{MattinglyStevenson} to the experimental measurements of the
$R$-ratio above the thresholds for the production of $c\overline{c}$
bound states provide the empirical constraint:  $\alpha_{R}({\sqrt s}
=5.0~{\rm GeV})/\pi \simeq 0.08\pm 0.03.$ The prediction for the
effective coupling for the deep inelastic sum rules at the commensurate
momentum transfer $Q$ is then $\alpha_{g_1}(Q=12.33\pm 1.20~{\rm
GeV})/\pi \simeq \alpha_{\rm GLS}(Q=12.33\pm 1.20~{\rm GeV})/\pi \simeq
0.074\pm 0.026.$ Measurements of the Gross-Llewellyn Smith sum rule have
so far only been carried out at relatively small values of
$Q^2$;\cite{CCFRL1,CCFRL2} however, one can use the results of the
theoretical extrapolation\cite{KS} of the experimental data presented
in:\cite{CCFRQ} $ \alpha_{\rm GLS}^{\rm extrapol}(Q=12.25~{\rm
GeV})/\pi\simeq 0.093\pm 0.042.$ This range overlaps with the prediction
from the generalized Crewther relation.  It is clearly important to have
higher precision measurements to fully test this fundamental QCD
prediction.

\section{Commensurate Scale Relations and Fixed Points}

In general, one can write the relation between any two effective charges
at arbitrary scales $Q_A$ and $Q_B$ as a correction to the corresponding
relation obtained in a conformally invariant theory:
\begin{equation}
\alpha_A(Q_A) =
 C^{AB}[\alpha_B(Q_B)]
+\Psi_B(\alpha_B(Q_B))D^{AB}[\alpha_B(Q_B)]
\end{equation}
where
\begin{equation}
C^{AB}[\alpha_B(Q_B)] =
\alpha_B(Q_B) +
\sum_{n=1}^{\infty} c_n^{AB}\frac{\alpha_B^{n+1}(Q_B)}{\pi^n}
\end{equation}
is the functional relation when $\Psi_B[\alpha_B]=0$.  In fact, if
$\alpha_B$ approaches a fixed point $\bar\alpha_B$ where
$\Psi_B[\bar\alpha_B]=0$, then $\alpha_A$ tends to a fixed point given
by
\begin{equation}
\alpha_A \to \bar\alpha_A = C_{AB}[\bar\alpha_B].
\label{eq:am}
\end{equation}
The commensurate scale relation for observables $A$ and $B$ has a
similar form, but in this case the relative scales
$\Lambda_{BA}=Q_B/Q_A$ are fixed such that the non-conformal term
$D^{AB}$ is zero.  Thus the commensurate scale relation $\alpha_A(Q_A) =
C^{AB}[\alpha_B(\Lambda_{BA}Q_A)]$ at general commensurate scales is
also the relation connecting the values of the fixed points for any two
effective charges or schemes.  Furthermore, as $\Psi\rightarrow 0$, the
ratio of commensurate scales $Q^2_A/Q^2_B$ becomes the ratio of fixed
point scales $\bar Q^2_A/\bar Q^2_B$ as one approaches the fixed point
regime.

\section{Implementation of $\alpha_V$ Scheme}
\unboldmath

The effective charge $\alpha_V(Q)$ provides a physically-based
alternative to the usual modified minimal subtraction
($\overline{\mbox{MS}}$) scheme.  All virtual corrections
due to fermion pairs are incorporated in $\alpha_V$ through loop diagrams
 which depend on the physical mass
thresholds.  When continued to time-like momenta, the coupling has the
correct analytic dependence dictated by the production thresholds in the
crossed channel.  Since $\alpha_V$ incorporates quark mass effects
exactly, it avoids the problem of explicitly computing and resumming
quark mass corrections which are related to the running of the coupling.
Thus the effective number of flavors $N_F(Q/m)$ is an analytic function
of the scale $Q$ and the quark masses $m$.  The effects of finite quark
mass corrections on the running of the strong coupling were first
considered by De R{\' u}jula and Georgi\cite{derujula} within the
momentum subtraction schemes (MOM) (see also
references\cite{Georgi_Politzer,Ross,shirkov,chyla}).  The two-loop
calculation was first done by Yoshino and Hagiwara\cite{yh} in the
MOM-scheme using Landau gauge and was also  recently 
calculated by Jegerlehner and
Tarasov\cite{jt} using background field gauge.

One important advantage of the physical charge approach is its inherent
gauge invariance to all orders in perturbation theory.  This feature is
not manifest in massive $\beta$-functions defined in non-physical
schemes such as the MOM schemes.  A second, more practical, advantage is
the automatic decoupling of heavy quarks according to the
Appelquist-Carazzone theorem.\cite{ac}

The specification of the coupling and renormalization scheme also
depends on the definition of the quark mass.  In contrast to QED where
the on-shell mass provides a natural definition of lepton masses, an
on-shell definition for quark masses is complicated by the confinement
property of QCD.  For a physical charge it is natural to use the pole
mass $m$ which has the advantage of being scheme and
renormalization-scale invariant as well as giving explicit decoupling.

In a recent paper\cite{bmr} with M.~Melles we have presented a two-loop
analytic extension of the $\alpha_V$-scheme based on previous
results.\cite{melles98} The mass effects are in principle treated
exactly to two-loop order and are only limited in practice by the
uncertainties from numerical integration.  The desired features of gauge
invariance and decoupling are manifest in the form of the two-loop
Gell-Mann Low function.  Strong consistency checks of the results are
performed by comparing the Abelian limit to the well known QED results
in the on-shell scheme.  In addition, the massless as well as the
decoupling limit are reproduced exactly, and the two-loop Gell-Mann Low
function is shown to be renormalization scale ($\mu$) independent.

\begin{figure}[htb]
\center
\epsfig{file=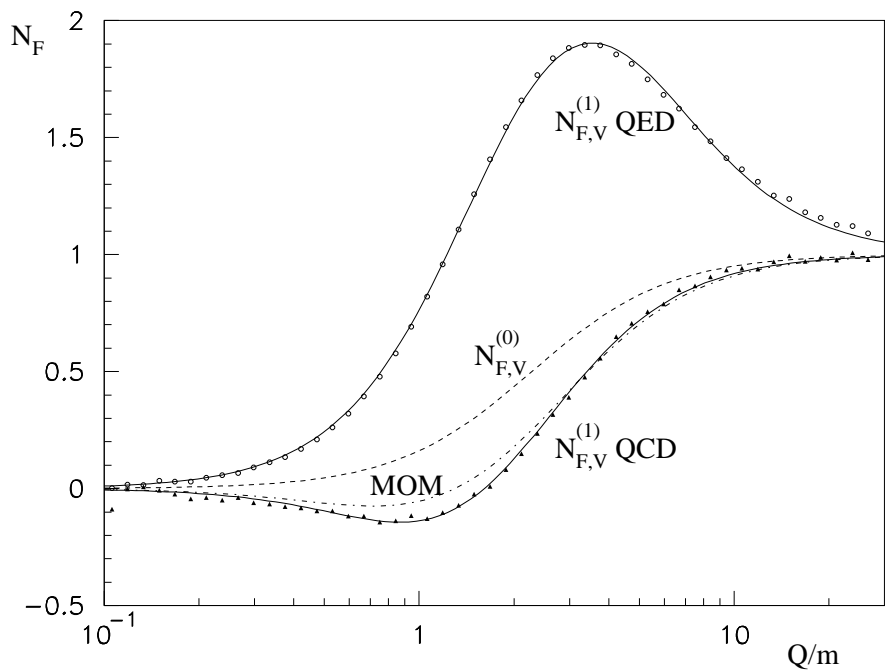,width=12cm}
\caption{The numerical results for the gauge-invariant $N_{F,V}^{(1)}$ in QED
(open circles) and QCD (triangles) with the parameterizations
superimposed respectively.  The dashed
line shows the parameterization of the one-loop $N_{F,V}^{(0)}$ function.
For comparison the gauge dependent two-loop result obtained in MOM schemes
(dash-dot) \protect\cite{yh,jt} is also shown.
 At large $Q/m$ the theory becomes
effectively massless, and both schemes agree as expected.  The figure also
illustrates the decoupling of heavy quarks at small $Q/m$.}
\label{fig:nfV}
\end{figure}

The results of the numerical calculation of $N_{F,V}^{(1)}$ in the
$V$-scheme for QCD and QED are shown in Fig.~\ref{fig:nfV}.  The
decoupling of heavy quarks becomes manifest at small $Q/m$, and the
massless limit is attained for large $Q/m$.  The QCD form actually
becomes negative at moderate values of $Q/m$, a novel feature of the
anti-screening non-Abelian contributions.  This property is also present
in the (gauge dependent) MOM results.  In contrast, in Abelian QED the
two-loop contribution to the effective number of flavors becomes larger
than one at intermediate values of $Q/m$.  The figure also displays the
one-loop contribution $N^{(0)}_{F,V} \left( Q/m \right)$ which
monotonically interpolates between the decoupling and massless limits.
The solid curves displayed in Fig.~\ref{fig:nfV} shows simple
parameterizations of $N_{F,V}^{(1)}$ which can be used to get a simple
representation of the numerical results.

By employing the commensurate scale relations, other physical observables
can be expressed in terms of the analytic coupling $\alpha_V$ without
scale or scheme ambiguity.  This way the quark mass threshold effects in
the running of the coupling are taken into account by utilizing the mass
dependence of the physical $\alpha_V$ scheme.  In effect, quark
thresholds are treated analytically to all orders in $m^2/Q^2$; {\it
i.e.}, the evolution of the physical $\alpha_V$ coupling in the
intermediate regions reflects the actual mass dependence of a physical
effective charge and the analytic properties of particle production.
Furthermore, the definiteness of the dependence in the quark masses
automatically constrains the scale $Q$ in the argument of the coupling.
There is thus no scale ambiguity in perturbative expansions in
$\alpha_V$.

The use of $\alpha_V$ and related physically defined effective charges
such as the plaquette charge $\alpha_P$ (to NLO the effective charge
defined from the (1,1) plaquette, $\alpha_P$ is the same as $\alpha_V$)
as expansion parameters has been found to be valuable in lattice gauge
theory, greatly increasing the convergence of perturbative expansions
relative to those using the bare lattice coupling.\cite{LepageMackenzie}
Recent lattice calculations of the $\Upsilon$- spectrum\cite{Davies}
have been used with BLM scale-fixing to determine a NLO normalization of
the static heavy quark potential:  $ \alpha_V^{(3)}(8.2 \GeV) =
0.196(3)$ where the effective number of light flavors is $N_F = 3$.  The
corresponding modified minimal subtraction coupling evolved to the $Z$
mass and five flavors is $ \alpha_{\overline{MS}}^{(5)}(M_Z) =
0.1174(24)$.  Thus a high precision value for $\alpha_V(Q^2)$ at a
specific scale is available from lattice gauge theory.  Predictions for
other QCD observables can be directly referenced to this value without
the scale or scheme ambiguities, thus greatly increasing the precision
of QCD tests.

One can also use $\alpha_V$ to characterize the coupling which appears
in the hard scattering contributions of exclusive process amplitudes at
large momentum transfer, such as elastic hadronic form factors, the
photon-to-pion transition form factor at large momentum
transfer\cite{BLM,BJPR} and exclusive weak decays of heavy
hadrons.\cite{Henley} Each gluon propagator with four-momentum $k^\mu$
in the hard-scattering quark-gluon scattering amplitude $T_H$ can be
associated with the coupling $\alpha_V(k^2)$ since the gluon exchange
propagators closely resembles the interactions encoded in the effective
potential $V(Q^2)$.  [In Abelian theory this is exact.] Commensurate
scale relations can then be established which connect the
hard-scattering subprocess amplitudes which control exclusive processes
to other QCD observables.

We can anticipate that eventually nonperturbative methods such as
lattice gauge theory or discretized light-cone quantization will provide
a complete form for the heavy quark potential in QCD.  It is
reasonable to assume that $\alpha_V(Q)$ will not diverge at small
space-like momenta.  One possibility is that $\alpha_V$ stays relatively
constant $\alpha_V(Q) \simeq 0.4$ at low momenta, consistent with
fixed-point behavior.  There is, in fact, empirical evidence for
freezing of the $\alpha_V$ coupling from the observed systematic
dimensional scaling behavior of exclusive reactions.\cite{BJPR} If this
is in fact the case, then the range of QCD predictions can be extended
to quite low momentum scales, a regime normally avoided because of the
apparent singular structure of perturbative extrapolations.

There are a number of other advantages of the $V$-scheme:
\begin{enumerate}
\item
Perturbative expansions in $\alpha_V$ with the scale set by the momentum
transfer cannot have any $\beta$-function dependence in their
coefficients since all running coupling effects are already summed into
the definition of the potential.  Since coefficients involving $\beta_0$
cannot occur in an expansions in $\alpha_V$, the divergent infrared
renormalon series of the form $\alpha^n_V\beta_0^n n!$ cannot occur.
The general convergence properties of the scale $Q^*$ as an expansion in
$\alpha_V$ is not known.\cite{Mueller}

\item
The effective coupling $\alpha_V(Q^2)$ incorporates virtual
contributions with finite fermion masses.  When continued
to time-like momenta, the coupling has the correct analytic dependence
dictated by the production thresholds in the $t$ channel.  Since
$\alpha_V$ incorporates quark mass effects exactly, it avoids the
problem of explicitly computing and resumming quark mass corrections.

\item
The $\alpha_V$ coupling is the natural expansion parameter for
processes involving non-relativistic momenta, such as heavy quark
production at threshold where the Coulomb interactions, which are
enhanced at low relative velocity $v$ as $\pi \alpha_V/v$, need to be
re-summed.\cite{Voloshin,Hoang,Fadin} The effective Hamiltonian for
nonrelativistic QCD is thus most naturally written in $\alpha_V$ scheme.
The threshold corrections to heavy quark production in $e^+ e^-$
annihilation depend on $\alpha_V$ at specific scales $Q^*$.  Two
distinct ranges of scales arise as arguments of $\alpha_V$ near
threshold:  the relative momentum of the quarks governing the soft gluon
exchange responsible for the Coulomb potential, and a high momentum
scale, induced by hard gluon exchange, approximately equal to twice the
quark mass for the corrections.\cite{Hoang} One thus can use threshold
production to obtain a direct determination of $\alpha_V$ even at low
scales.  The corresponding QED results for $\tau$ pair production allow
for a measurement of the magnetic moment of the $\tau$ and could be
tested at a future $\tau$-charm factory.\cite{Voloshin,Hoang}

\end{enumerate}

We also note that computations in different sectors of the Standard
Model have been traditionally carried out using different
renormalization schemes.  However, in a grand unified theory, the forces
between all of the particles in the fundamental representation should
become universal above the grand unification scale.  Thus it is natural
to use $\alpha_V$ as the effective charge for all sectors of a grand
unified theory, rather than in a convention-dependent coupling such as
$\alpha_{\overline {\tiny MS}}$.

\section{The Analytic Extension of the $\overline{\mbox{MS}}$ Scheme}

The standard $\overline{\mbox{MS}}$ scheme is not an analytic function
of the renormalization scale at heavy quark thresholds; in the running
of the coupling the quarks are taken as massless, and at each quark
threshold the value of $N_F$ which appears in the $\beta$ function is
incremented.  However, one can use the commensurate scale relation
between $\alpha_V(Q)$ and the conventional $\overline{\mbox{MS}}$ coupling
to define an extended $\overline{\mbox{MS}}$ scheme which is continuous
and analytic at any scale.  The new modified scheme inherits most of the
good properties of the $\alpha_V$ scheme, including its correct analytic
properties as a function of the quark masses and its unambiguous scale
fixing.\cite{bgmr} The modified coupling is defined as
\begin{equation}
\widetilde {\alpha}_{\overline{\mbox{\tiny MS}}}(Q)
= \alpha_V(Q^*,m) + \frac{2N_C}{3} {\alpha_V^2(Q^{*},m)\over\pi} +
\cdots ,
\label{alpmsbar2}
\end {equation}
for all perturbative scales $Q$, where the LO commensurate scale is given by $Q^* =
Q\exp\left(5/6\right)$ and $m$ are the pole-masses.  This equation not
only provides an analytic extension of the $\overline{\mbox{MS}}$ and
similar schemes, but it also ties down the renormalization scale to the
masses of the quarks as they enter into the virtual corrections to $\alpha_V$.

The coefficients in the perturbation expansion have their conformal
values, \ie, the same coefficients would occur even if the theory had
been conformally invariant with $\beta=0$.  The coefficient $2 N_C/3$ in
the NLO coefficient is a feature of the non-Abelian couplings of QCD;
the same coefficient occurs even if the theory were conformally
invariant with $\beta_0=0.$

The modified scheme \amst\ provides an analytic interpolation of
conventional $\overline{\mbox{MS}}$ expressions by utilizing the mass
dependence of the physical \av\ scheme.  In effect, quark thresholds are
treated analytically to all orders in $m^2/Q^2$; \ie, the evolution of
the analytically extended coupling in the intermediate regions reflects
the actual mass dependence of a physical effective charge and the
analytic properties of particle production.    Furthermore, the
definiteness of the dependence in the quark masses automatically
constrains the renormalization scale.  There is thus no scale ambiguity
in perturbative expansions in \av\ or \amst.

In leading order the effective number of flavors in the modified scheme
\amst\ is given to a very good approximation by the simple form\cite{bgmr}
\begin{equation}
\widetilde {N}_{F,\overline{\mbox{\tiny MS}}}^{(0)}\left(\frac{m^2}{Q^2}\right)
\cong \left(1 + {5m^2 \over {Q^2\exp({5\over 3})}} \right)^{-1}
\cong \left( 1 + {m^2 \over Q^2} \right)^{-1}.
\end{equation}
Thus the contribution from one flavor is $\simeq 0.5$ when
the scale $Q$ equals the quark mass $m$.  The standard procedure
of matching $\alpha_{\overline{\mbox{\tiny MS}}}(\mu)$ at the quark
masses serves as a zeroth-order approximation to the continuous $N_F$.

\begin{figure}[htb]
\begin{center}
\leavevmode
\epsfxsize=4in
\epsfbox{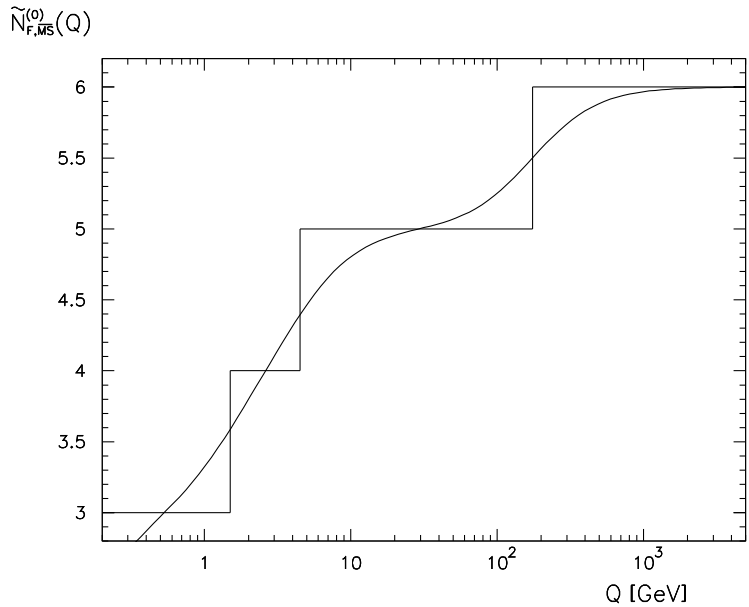}
\end{center}
\caption[*]{The continuous
$\widetilde {N}_{F,\overline{\mbox{\tiny MS}}}^{(0)}$ in the analytic
extension of the $\overline{\mbox{MS}}$ scheme as a
function of the physical scale $Q$.  (For reference the
continuous $N_F$ is also compared with
the conventional procedure of taking $N_F$ to be a step-function at the
quark-mass thresholds.)}
\label{fig:nfsum}
\end{figure}

Adding all flavors together gives the total $\widetilde
{N}_{F,\overline{\mbox{\tiny MS}}}^{(0)}(Q)$ which is shown in
Fig.~\ref{fig:nfsum}.  For reference, the continuous $N_F$ is also
compared with the conventional procedure of taking $N_F$ to be a
step-function at the quark-mass thresholds.  The figure shows clearly
that there are hardly any plateaus at all for the continuous $\widetilde
{N}_{F,\overline{\mbox{\tiny MS}}}^{(0)}(Q)$ in between the quark
masses.  Thus there is really no scale below 1 TeV where $\widetilde
{N}_{F,\overline{\mbox{\tiny MS}}}^{(0)}(Q)$ can be approximated by a
constant; for all $Q$ below 1 TeV there is always one quark with mass
$m$ such that $m^2 \ll Q^2$ or $Q^2 \gg m^2$ is not true.  We also note
that if one would use any other scale than the BLM-scale for $\widetilde
{N}_{F,\overline{\mbox{\tiny MS}}}^{(0)}(Q)$, the result would be to
increase the difference between the analytic $N_F$ and the standard
procedure of using the step-function at the quark-mass thresholds.

\begin{figure}[htb]
\begin{center}
\leavevmode
\epsfxsize=4in
\epsfbox{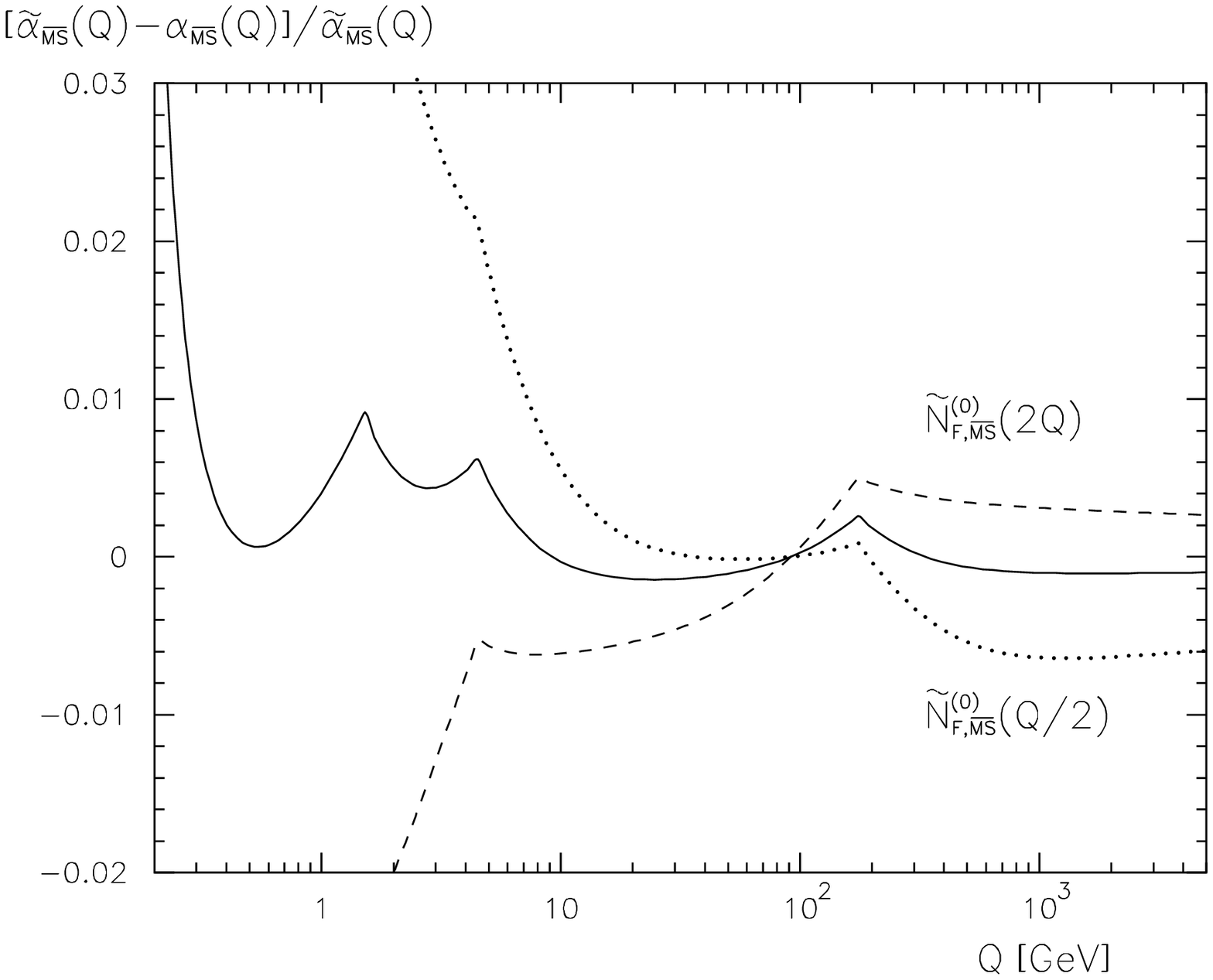}
\end{center}
\caption[*]{The solid curve shows the relative difference between the
solutions to
the 1-loop renormalization group equation using continuous $N_F$,
$\widetilde{\alpha}_{\overline{\mbox{\tiny MS}}}(Q)$, and conventional discrete
theta-function thresholds, $\alpha_{\overline{\mbox{\tiny MS}}}(Q)$.
The dashed (dotted) curves shows the same quantity but using the scale $2Q$
($Q/2$)
in $\widetilde {N}_{F,\overline{\mbox{\tiny MS}}}^{(0)}$.  The solutions
have been
obtained numerically starting from the world average\cite{Burrows}\
$\alpha_{\overline{\mbox{\tiny MS}}}(M_Z) = 0.118$.}
\label{fig:adiff}
\end{figure}

Figure~\ref{fig:adiff} shows the relative difference between the two
different solutions of the 1-loop renormalization group equation,
\ie\ $(\widetilde{\alpha}_{\overline{\mbox{\tiny MS}}}(Q)-
           {\alpha}_{\overline{\mbox{\tiny MS}}}(Q) )/
           \widetilde{\alpha}_{\overline{\mbox{\tiny MS}}}(Q)$.
The solutions have been obtained numerically starting from the
world average\cite{Burrows}
$\alpha_{\overline{\mbox{\tiny MS}}}(M_Z) = 0.118$.
The figure shows that
taking the quark masses into account in the running leads to
effects of the order of one percent which are most especially
pronounced near thresholds.

The extension of the $\overline{\mbox{MS}}$-scheme provides a coupling
which is an analytic function of both the scale and the quark masses.
The modified coupling $\widetilde {\alpha}_{\overline{\mbox{\tiny
MS}}}(Q)$ inherits most of the good properties of the $\alpha_V$ scheme,
including its correct analytic properties as a function of the quark
masses and its unambiguous scale fixing\cite{bgmr}.  However, the
conformal coefficients in the commensurate scale relation between the
$\alpha_V$ and $\overline{\mbox{MS}}$ schemes does not preserve one of
the defining criterion of the potential expressed in the bare charge,
namely the non-occurrence of color factors corresponding to an iteration
of the potential.  This is probably an effect of the breaking of
conformal invariance by the $\overline{\mbox{MS}}$ scheme.  The breaking
of conformal symmetry has also been observed when dimensional
regularization is used as a factorization scheme in both
exclusive\cite{Frishman, Muller} and inclusive\cite{Blumlein} reactions.
Thus, it does not turn out to be possible to extend the modified scheme
${\widetilde \alpha}_{\overline{\mbox{\tiny MS}}}$ beyond leading order
without running into an intrinsic contradiction with conformal symmetry.

\section{ Application to Hadronic $Z$ Decay }\label{sec:appl}

This section shows how the analytic $\alpha_V$-scheme can be used to
calculate the non-singlet hadronic width of the Z-boson, including
finite quark mass corrections from the running of the
coupling\cite{bmr}.  The results are compared with the standard
treatment in the $\overline{\mbox{MS}}$ scheme where finite quark mass
effects are calculated as higher twist corrections.

\begin{figure}[htbp]
\begin{center}
\mbox{\epsfig{figure=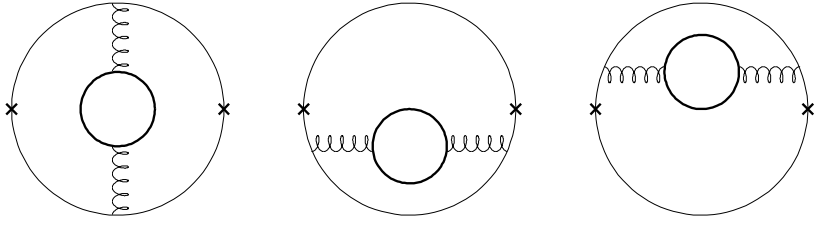,width=12cm}}
\end{center}
\caption[*]{The ``double bubble" diagrams.  The crosses represent
the external electro weak current, the thin line is a massless quark
and the thick line is a massive quark.}
\label{fig:doublebubble}
\end{figure}

The finite quark mass effects which are of interest are in leading order
given by the ``double bubble" diagrams, which are shown in
Fig.~\ref{fig:doublebubble}, where the outer quark loop which couples to
the weak current is considered massless and the inner quark loop is
massive.  These correction have been calculated in the
$\overline{\mbox{MS}}$ scheme as expansions in $m_q^2/s$\cite{Hoang} and
$s/m_Q^2$\cite{Chetyrkin} for light and heavy quarks, respectively,
whereas they have been calculated numerically\cite{Soper_Surguladze}.
In addition the $\alpha_{\mbox{\scriptsize{s}}}^3$ correction due to
heavy quarks has been calculated as an expansion in $s/m_Q^2$
in\cite{L_R_V_NPB}.  Other types of mass corrections, such as the
double-triangle graphs where the external current is electroweak, are
not taken into account.

The non-singlet hadronic width of
a hypothetical Z-boson with mass $\sqrt{s}$ is given by
\begin{equation}
\Gamma_{had}^{NS}(s)=\frac{G_Fs^{3/2}}{2\pi\sqrt{2}}
\sum_{q}\{(g_V^{q})^2+(g_A^{q})^2\}
\left[1+\frac{3}{4}C_F\frac{\alpha_{\Gamma}^{NS}(s)}{\pi}\right] \;,
\end{equation}
where $\alpha_{\Gamma}^{NS}(s)$ is the effective charge\cite{Grunberg} which
contains all QCD corrections.

In the following, the next-to-leading order expressions for the
effective charge $\alpha_{\Gamma}^{NS}(s)$ in the $\overline{\mbox{MS}}$
and V schemes will be compared for arbitrary $s$ using next-to-leading
order evolution starting from the physical mass $\sqrt{s}=M_Z$ which is
used as normalization condition.

In the $\overline{\mbox{MS}}$ scheme the
effective charge $\alpha_{\Gamma}^{NS}(s)$ is to next-to-leading order
 given by
\begin{eqnarray} \label{eq:agms}
\alpha_{\Gamma}^{NS}(s) & = &
\alpha_{\overline{\mbox{\tiny MS}}}^{(N_L)}(\mu)  \\
&& +\left[r_{1,\overline{\mbox{\tiny MS}}}(\mu)+
\sum_{q=1}^{N_L}F_1\left(\frac{m_q^2}{s}\right)+
\sum_{Q=N_L+1}^{6}G_1\left(\frac{s}{m_Q^2}\right)\right]
\frac{\left(\alpha_{\overline{\mbox{\tiny MS}}}^{(N_L)}(\mu)\right)^2}{\pi}
\nonumber
\end{eqnarray}
where the coefficient  $r_1$  is given by,
\begin{eqnarray*}
r_{1,\overline{\mbox{\tiny MS}}}(\mu=\sqrt{s})
& = & -\frac{1}{8}C_F+\frac{1}{12}N_C+
\left(\frac{11}{4}-2\zeta_3\right)\beta_0
= 1.986-0.115N_F
\end{eqnarray*}
(with $\beta_0=\psi_V^{(0)}(m=0)$  ) and
the functions $F_1$ and $G_1$ are the effects of non-zero quark masses
for light and heavy quarks, respectively.  The expansions of the
 finite quark mass corrections are given by
\begin{eqnarray}
F_1\left(\frac{m^2}{s}\right) & = &
\left(\frac{m^2}{s}\right)^2\left[\frac{13}{3}
-4\zeta_3-\ln\left(\frac{m^2}{s}\right)\right]
\nonumber \\ &&
+ \left(\frac{m^2}{s}\right)^3\left[\frac{136}{243}+\frac{16}{27}\zeta_2
+\frac{56}{81}\ln\left(\frac{m^2}{s}\right)
-\frac{8}{27}\ln^2\left(\frac{m^2}{s}\right)\right]
\\
G_1\left(\frac{s}{m^2}\right) & = &
\frac{s}{m^2}\left[ \frac{44}{675}
-\frac{2}{135}\ln\left(\frac{s}{m^2}\right)\right] \nonumber \\
& & +\left(\frac{s}{m^2}\right)^2\left[-\frac{1303}{1058400}
+\frac{1}{2520}\ln\left(\frac{s}{m^2}\right)\right]
\end{eqnarray}
which are good to within a few percent for $m_q^2/s<0.25$
and $s/m_Q^2<4$ respectively.
In addition the relation,\cite{Soper_Surguladze}
\begin{equation}
F\left(\frac{m^2}{s}\right) =
G\left(\frac{m^2}{s}\right) + \frac{1}{6}\ln\left(\frac{m^2}{s}\right)
-\left(-\frac{11}{12}+\frac{2}{3}\zeta_3 \right)
\end{equation}
is used to obtain $F_1$ in the interval $0.25 < m^2/s < 1$ where the expansion
of $F_1$ given above breaks down.

The number of light flavors $N_L$ in the $\overline{\mbox{MS}}$ scheme
is a function of the renormalization scale $\mu$.  In the following it
is assumed that the matching of the different effective theories with
different number of massless quarks is done at the quark masses.  In
other words a quark with mass $m<\mu$ is considered as light whereas a
quark with mass $m>\mu$ is considered as heavy.  In addition the
$\overline{\mbox{MS}}$ quark masses are used.  The dependence on the
matching scale can be made arbitrarily small by calculating the matching
condition to high enough order.  However this does not mean that the
finite quark mass effects are taken into account.  The only way to
include these mass effects in the ordinary $\overline{\mbox{MS}}$
treatment is by making a higher twist expansion to all orders in
$m^2/Q^2$ and $Q^2/m^2$ for light and heavy quarks respectively, {\it
i.e.} the functions $F$ and $G$ given above.  In the following
comparison $\mu=\sqrt{s}$ is used and the matching is done at the quark
masses.

\begin{figure}[htb]
\begin{center}
\mbox{\epsfig{figure=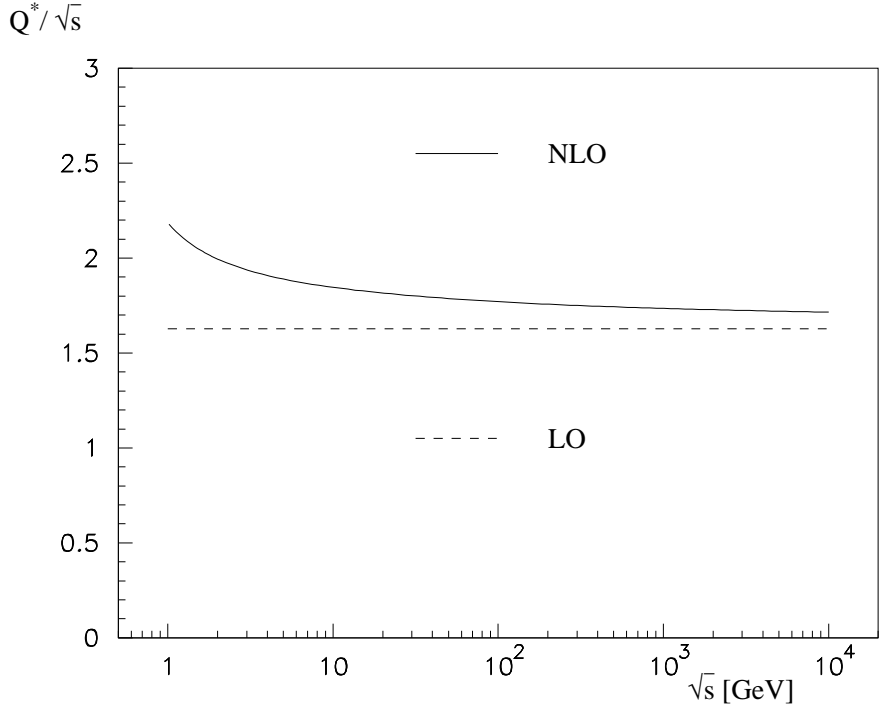,width=12cm}}
\end{center}
\caption[*]{The ratio of the commensurate scale $Q^*$ to $\sqrt{s}$
between the non-singlet width of the Z-boson and the heavy quark potential
as a function of $\sqrt{s}$ in next-to-leading (solid)
and leading (dashed) order.}
\label{fig:qratv}
\end{figure}

The commensurate scale relation between
$\alpha_{\Gamma}^{NS} $ and $\alpha_V$ is given by,\cite{bmr}
\begin{eqnarray}
\alpha_{\Gamma}^{NS}(\sqrt{s}) & = &
\alpha_V(Q^*,m_i) +
\left(-\frac{1}{8}C_F+\frac{3}{4}N_C\right) {\alpha_V^2(Q^*,m_i) \over \pi} +
\nonumber \\ &&
\left[-\frac{23}{32}C_F^2+\frac{21}{16}C_F N_C+
\left(-\frac{16\pi^2-\pi^4}{64}-\frac{7}{24}\right)N_C^2
 \right]{\alpha_V^3(Q^*,m_i) \over \pi^2} ,
\nonumber \\ & = &
\alpha_V(Q^*,m_i) +
2.083 {\alpha_V^2(Q^*,m_i) \over \pi}-7.161 {\alpha_V^3(Q^*,m_i) \over \pi^2} ,
\label{eq:agvfull}
\end{eqnarray}
where $Q^*$ is the commensurate scale and $m_i$ are the pole masses.
To leading order $Q^*$ is given
by $$Q^* = \exp\left(-\frac{23}{12}+2\zeta_3\right)\sqrt{s} = 1.628 \sqrt{s}$$
whereas to next-to-leading order it is given by
\begin{eqnarray}\label{eq:csr}
\frac{Q^*}{\sqrt s} & = & \exp\left\{-\frac{23}{12}+2\zeta_3
+\frac{\displaystyle
  \left[
  a_1 \psi_V^{(0)}(Q^*,m_i)+ a_2 \left(\psi_V^{(0)}(Q^*,m_i)\right)^2
  \right] \frac{\alpha_V(Q^*,m_i)}{\pi}  }
{\displaystyle\psi_V^{(0)}(Q^*,m_i) +  \psi_V^{(1)}(Q^*,m_i)\frac{\alpha_V(Q^*,m_i)}{\pi}}
\right\}\ . \nonumber
\end{eqnarray}
where
\begin{eqnarray}
a_1&=& \left(\frac{25}{16}-7\zeta_3-10\zeta_5\right)C_F
  +\left(-\frac{5}{3}+\frac{7}{12}\zeta_3+\frac{5}{3}\zeta_5\right)N_C = 1.765
\nonumber \\
a_2&=& -\frac{119}{144}-\frac{14}{3}\zeta_3+4\zeta_3^2+\frac{\pi^2}{12} = 0.166.
\end{eqnarray}
The resulting commensurate scale $Q^*$ is shown in Fig.~\ref{fig:qratv}
where it is also compared with the leading order scale.  As can be seen
from the figure the next-to-leading order correction to
the commensurate scale is small.
The general convergence properties of the scale $Q^*$ as an
expansion in $\alpha_V$ is not known.\cite{Mueller}

It should be noted that the scale $Q^*$ is only known to next-to-leading order.
Similarly the evolution equation for $\alpha_V(Q,m_i)$ is only known
to next-to-leading order.
Therefore one can only consistently use the next-to-leading order result
when comparing with the treatment
of finite quark mass effects $\overline{\mbox{MS}}$ scheme, {\it i.e.}
\begin{eqnarray}\label{eq:agv}
\alpha_{\Gamma}^{NS}(\sqrt{s}) & = &
\alpha_V(Q^*,m_i) +
\left(-\frac{1}{8}C_F+\frac{3}{4}N_C\right) {\alpha_V^2(Q^*,m_i) \over \pi}
\end{eqnarray}
where the scale $Q^*$ should be the leading order result for consistency.

\begin{figure}[htb]
\begin{center}
\mbox{\epsfig{figure=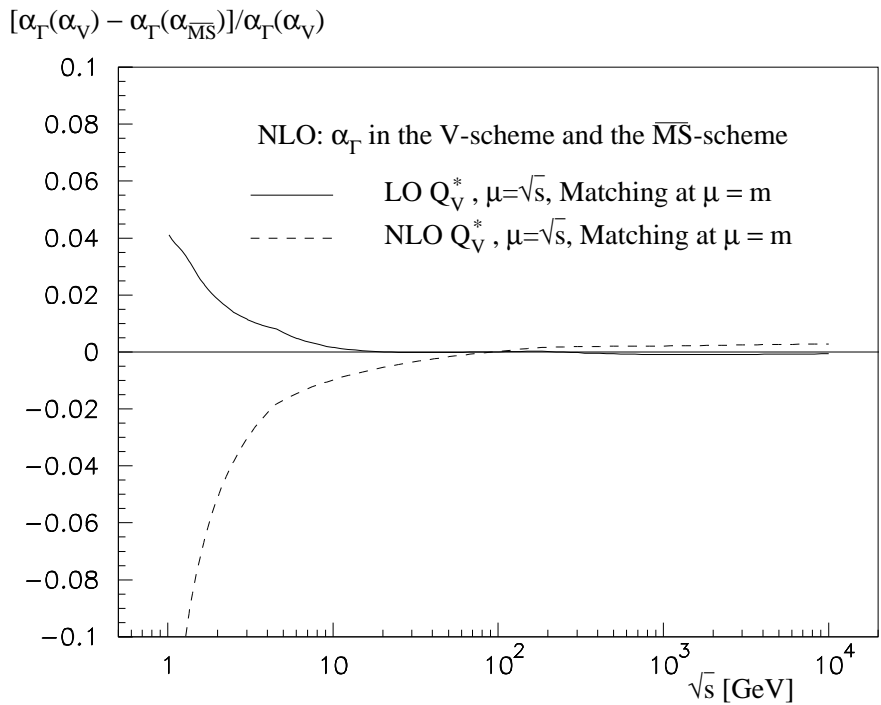,width=12cm}}
\end{center}
\caption[*]{ The relative difference between the
next-to-leading order expressions for $\alpha_{\Gamma}^{NS}(\sqrt{s})$
in the $\overline{\mbox{MS}}$ and V schemes respectively using next-to-leading
order evolution.}
\label{fig:agdiff}
\end{figure}

Figure~\ref{fig:agdiff} shows the relative difference between the
next-to-leading order expressions for $\alpha_{\Gamma}^{NS}(\sqrt{s})$
in the $\overline{\mbox{MS}}$ and V schemes given by
Eqs.~(\ref{eq:agms}) and (\ref{eq:agv}) respectively.
The predictions have been
normalized to the same value at $\sqrt{s}=M_Z$
using $\alpha_{\overline{\mbox{\tiny MS}}}^{(5)}(M_Z)=0.118$ and then
evolved using next-to-leading order evolution in the respective schemes.

The comparison shown in Fig.~\ref{fig:agdiff} illustrates the relative
difference between the predictions for $\alpha_{\Gamma}^{NS}(\sqrt{s})$
in the $\overline{\mbox{MS}}$ and V schemes.  It has been shown\cite{bgmr}
that the different ways of including the finite quark mass effects is
smaller than $\sim 0.1\%$ by comparing the $\overline{\mbox{MS}}$ scheme
with the analytic extension of the same which properly takes into
account the flavor threshold effects analytically.  Therefore the
difference between the $\overline{\mbox{MS}}$ and V scheme predictions
for $\alpha_{\Gamma}^{NS}(\sqrt{s})$ can be attributed to the scheme
dependence.  This is illustrated by the fact that when using the
next-to-leading order approximation for the commensurate scale, instead
of the leading order one, the relative difference changes sign and even
becomes larger.  This sensitivity is a consequence of the scale
dependence of the coupling, especially at small scales where the
$\Psi$-function is large.  The proper inclusion of the finite quark mass
effects is verified by the smoothness of the curve.

\section{Application of Commensurate Scale Relations to the Hard QCD
\hfill\break Pomeron}

The observation of rapidly increasing structure functions in deep
inelastic scattering at small-$x_{bj}$ and the observation of rapidly
increasing diffractive processes such as $\gamma^* p \to \rho p$ at high
energies at HERA is in agreement with the expectations of the
BFKL\cite{BFKL} QCD high-energy limit.  The highest eigenvalue,
$\omega^{\max}$, of the LO BFKL equation\cite{BFKL} is
related to the intercept of the Pomeron which in turn governs the
high-energy asymptotics of the cross sections:  $\sigma \sim
s^{\alpha_{I \negthinspace P}-1} = s^{\omega^{\max}}$.  The BFKL Pomeron
intercept in LO turns out to be rather large:  $\alpha_{I \negthinspace
P} - 1 =\omega_L^{\max} = 12 \, \ln2 \, ( \alpha_S/\pi ) \simeq 0.55 $
for $\alpha_S=0.2$; hence, it is very important to know the
NLO corrections.

Recently the NLO corrections to the BFKL resummation of energy
logarithms were calculated\cite{FL,CC98} by employing the $\overline{\mbox{MS}}$
scheme  to
regulate the ultraviolet divergences with arbitrary scale setting.  The
NLO corrections to the highest eigenvalue of the BFKL equation turn out
to be negative and even larger than the LO contribution for $\alpha_s >
0.157$.  It is thus important to analyze the NLO BFKL resummation of
energy logarithms in physical renormalization schemes and apply the
BLM-CSR method.  In fact, as shown in a recent paper,\cite{BBFKL} the
reliability of QCD predictions for the intercept of the BFKL Pomeron at
NLO when evaluated using BLM scale setting\cite{BLM} within non-Abelian
physical schemes, such as the momentum space subtraction (MOM)
scheme\cite{Cel79,Pas80} or the $\Upsilon$-scheme based on $\Upsilon
\rightarrow ggg$ decay, is significantly improved compared to the
$\overline{\mbox{MS}}$-scheme.

The renormalization scale ambiguity problem can be resolved if one can
optimize the choice of scales and renormalization schemes according to
some sensible criteria.  In the BLM optimal scale setting,\cite{BLM} the
renormalization scales are chosen such that all 
effects  related to the QCD $\beta$-function are resummed into the running
couplings.  The coefficients of the perturbative series are thus
identical to the perturbative coefficients of the corresponding
conformally invariant theory with $\beta=0$.

In the present case one can show that within the V-scheme (or the
$\overline{\mbox{MS}}$-scheme) the BLM procedure does not change
significantly the value of the NLO coefficient $r(\nu)$.  This can be
understood since the V-scheme, as well as $\overline{\mbox{MS}}$-scheme,
are adjusted primarily to the case when in the LO there are dominant QED
(Abelian) type contributions, whereas in the BFKL case there are
important LO gluon-gluon (non-Abelian) interactions.  Thus one can
choose for the BFKL case the MOM-scheme\cite{Cel79,Pas80} or the
$\Upsilon$-scheme based on $\Upsilon \rightarrow ggg$ decay.

Adopting BLM scale setting, the NLO BFKL eigenvalue
in the MOM-scheme is
\begin{equation}
\omega_{BLM}^{MOM}(Q^{2},\nu) =
N_C \chi_{L} (\nu) \frac{\alpha_{MOM}(Q^{MOM \, 2}_{BLM})}{\pi}
\Biggl[1 +
r_{BLM}^{MOM} (\nu) \frac{\alpha_{MOM}(Q^{MOM \, 2}_{BLM})}{\pi} \Biggr] ,
\end{equation}
where
$ r_{BLM}^{MOM} (\nu) $ is given numerically in Table \ref{tab:2}
and the   BLM  scale is given by,
\begin{equation}
Q^{MOM \, 2}_{BLM} (\nu)
= Q^2 \exp \Biggl[ \frac
{1}{2}\chi_L (\nu) - \frac{5}{3} + 2 \biggl(1+\frac{2}{3} I \biggr) \Biggr] \ ,
 \nonumber
\label{qblm}
\end{equation}
where $I=-2\int^1_0 dx\, {\rm ln}(x)/[x^2-x+1] \sim 2.3439$.
At $\nu=0$ we
have $Q^{MOM \, 2}_{BLM} (0) = Q^2 \bigl( 4 \exp [2(1+2 I /3)-5/3] \bigr) \simeq
Q^2  \, 127$. Note that $Q^{MOM \, 2}_{BLM}(\nu)$ contains a large factor,
$  \exp [2(1+2 I /3)] \simeq 168$, which
reflects a large kinematic difference between MOM- and
$\overline{\mbox{MS}}$- schemes.\cite{Cel83,BLM}

\begin{figure}[htb]
\begin{center}
\leavevmode
{\epsfxsize=12cm\epsfysize=12cm\epsfbox{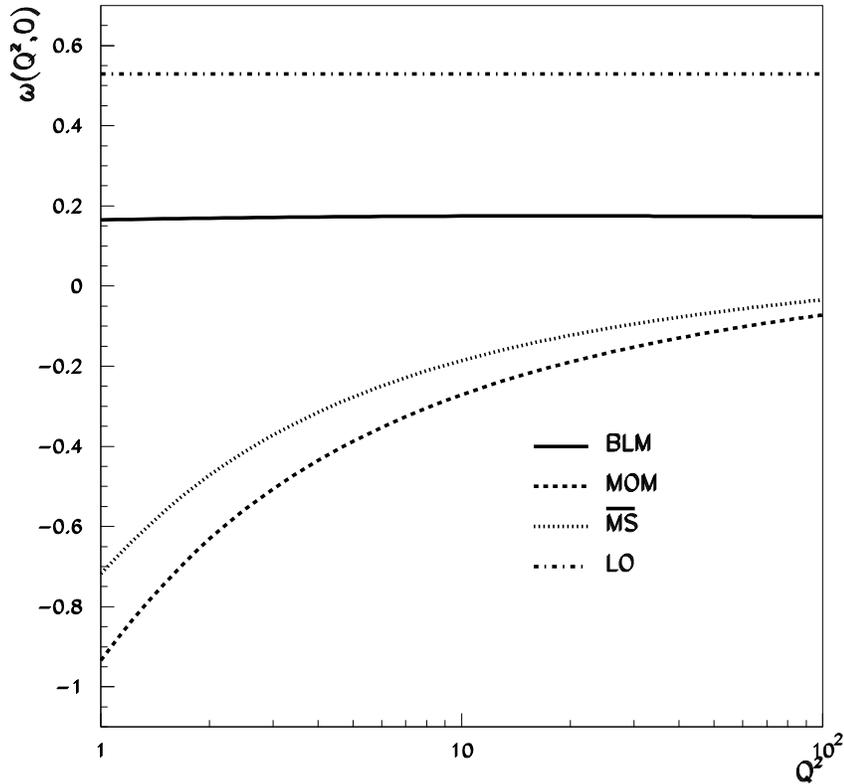}}
\end{center}
\caption[*]{$Q^2$-dependence of the BFKL Pomeron intercept in the NLO.
  BLM (in MOM-scheme) -- solid,
MOM-scheme (Yennie gauge: $\xi=3$) -- dashed,
$\overline{\mbox{MS}}$-scheme -- dotted. LO
BFKL ($\alpha_S=0.2$) -- dash-dotted.}
\label{fig:2}
\end{figure}

Figure \ref{fig:2} gives the result for the $Q^2$ dependence of the
eigenvalue of the NLO BFKL kernel using the QCD parameter $\Lambda = 0.1$
GeV which corresponds to $\alpha_S = 4 \pi / \bigl[ \beta_0
\ln(Q^2/\Lambda^2) \bigr] \simeq 0.2$ at $Q^2=15$ GeV$^2$.

One of the striking features of this analysis is that the NLO value for
the intercept of the BFKL Pomeron, improved by the BLM procedure, has a
very weak dependence on the gluon virtuality $Q^2$.  This agrees with
the conventional Regge-theory where one expects an universal intercept
of the Pomeron without any $Q^2$-dependence.  The minor $Q^2$-dependence
obtained, on one side, provides near insensitivity of the results to the
precise value of $\Lambda$, and, on the other side, leads to approximate
scale and conformal invariance.  Thus one may use conformal
symmetry\cite{Lipatov97,Lipatov86} for the continuation of the present
results to the case $t \neq 0$.

\begin{table}%
\begin{center}
\begin{tabular}{|c|c|c|c|c|c|}
\hline
\multicolumn{2}{|c|}{Scheme} & \multicolumn{1}{|c|}{$r_{BLM}(0)$} &
\multicolumn{3}{|c|}
{$\alpha_{I \negthinspace P}^{BLM} - 1 =\omega_{BLM}(Q^2,0)$} \\
\cline{4-6}
\multicolumn{2}{|c|}{}  & \multicolumn{1}{|c|}{$(N_F = 4)$} &
$Q^2=1$ GeV$^2$ & $Q^2=15$ GeV$^2$ & $Q^2=100$ GeV$^2$    \\
\hline\hline
M & $\xi=0$ & -13.05 & 0.134 & 0.155 & 0.157 \\
 \cline{2-6}
O & $\xi=1$ & -12.28 & 0.152 &  0.167 & 0.166 \\
\cline{2-6}
M & $\xi=3$ & -11.74 & 0.165 & 0.175 & 0.173 \\
\hline
\multicolumn{2}{|c|}{$\Upsilon$} & -14.01 & 0.133  & 0.146 & 0.146 \\
\hline
\end{tabular}
\end{center}
\caption{The NLO BFKL Pomeron intercept in the BLM scale setting
within non-Abelian physical schemes.}
\label{tab:2}
\end{table}

The NLO corrections to the BFKL equation for the QCD Pomeron thus become
controllable and meaningful provided one uses physical renormalization
scales and schemes relevant to non-Abelian gauge theory.  BLM scale
setting automatically sets the appropriate physical renormalization
scale by absorbing the non-conformal $\beta$-dependent coefficients.
These results open new windows for applications of NLO BFKL resummation
to high-energy phenomenology. and constitutes a first step towards a
more complete understanding of the NLO corrections to the structure
functions in the Regge limit.  An important issue raised by
R.~Thorne\cite{Thorne} is the factorization scheme dependence of the
result.  Ideally one should make the analysis employing a physical
factorization scheme.\cite{Grunberg,catani}

Recently the $L3$ collaboration at LEP has presented new results for
the virtual photon cross section $\sigma(\gamma^*(Q_A) \gamma^*(Q_b) \to
{\rm hadrons})$ using double tagged $e^+ e^- \to e^+ e^- {\rm hadrons}.$
This process provides a remarkably clean possible test of the
perturbative QCD pomeron since there are no initial
hadrons.\cite{bhsprl} The calculation of $\sigma (\gamma^* \gamma^*)$ is
discussed in detail in references.\cite{bhsprl} We note here some
important features:

i) For large virtualities, $\sigma (\gamma^* \gamma^*)$ the longitudinal
cross section $\sigma_{LL}$ dominates and scales like $1/Q^2$, where
$Q^2 \sim {\mbox {\rm max}} \{ Q_A^2, Q_B^2\} $. This is characteristic
of the perturbative QCD prediction.  Models based on Regge factorization
(which work well in the soft-interaction regime dominating $\gamma \,
\gamma$ scattering near the mass shell) would predict a higher power in
$1/Q$.

ii) $\sigma (\gamma^* \gamma^*)$ is affected by logarithmic corrections
in the energy $s$ to all orders in $\alpha_s$.  As a result of the BFKL
summation of these contributions, the cross section rises like a power
in $s$, $\sigma \propto s^\lambda$.  The Born approximation to this
result; that is, the ${\cal O} (\alpha_s^2) $ contribution,
corresponding to single gluon exchange gives a constant cross section,
$\sigma_{\rm Born} \propto s^0$.  A fit to photon-photon sub-energy
dependence measured by L3 at $\sqrt s_{e^+e^-} = 91~{\rm GeV}$ and
$\VEV{Q_A^2}=\VEV{Q_B^2} = 3.5~{\rm GeV}^2$ gives $\alpha_P-1 = 0.28 \pm
0.05$.  The L3 data at $\sqrt s_{e^+e^-} = 183~{\rm GeV}$ and
$\VEV{Q_A^2} = \VEV{Q_B^2} = 14~{\rm GeV}^2,$ gives $\alpha_P-1 = 0.40
\pm 0.07$ which shows a rise of the virtual photon cross section much
stronger than single gluon or soft pomeron exchange, but it is
compatible with the expectations from the NLO scale- and scheme-fixed
BFKL predictions.  It will be crucial to measure the $Q_A^2$ and $Q_B^2$
scaling and polarization dependence and compare with the detailed
predictions of PQCD.\cite{bhsprl}

\section{Conclusions}

Commensurate scale relations have a number of attractive properties:
\begin{enumerate}
\item
The ratio of physical scales $Q_A/Q_B$ which
appears in commensurate scale relations reflects the relative position
of physical thresholds, \ie\, quark anti-quark pair production.
\item
The functional dependence and perturbative expansion of the CSR are
identical to those of a conformal scale-invariant theory where
$\beta_A(\alpha_A)=0$ and $\beta_B(\alpha_B)=0$.
\item
In the case of
theories approaching fixed-point behavior $\beta_A(\bar\alpha_A)=0$ and
$\beta_B(\bar\alpha_B)=0$, the commensurate scale relation relates both
the ratio of fixed point couplings $\bar\alpha_A/\bar\alpha_B$, and the
ratio of scales as the fixed point is approached.
\item
Commensurate
scale relations satisfy the Abelian correspondence principle
\cite{BrodskyHuet}; \ie\ the non-Abelian gauge theory prediction reduces
to Abelian theory for $N_C \to 0$ at fixed $ C_F\alpha_s$ and fixed
$N_F/C_F$.
\item
The perturbative expansion of a commensurate scale
relation has the same form as a conformal theory, and thus has no $n!$
renormalon growth arising from the $\beta$-function.  It is an
interesting conjecture whether the perturbative expansion relating
observables to observable are in fact free of all $n!$ growth.  The
generalized Crewther relation, where the commensurate relation's
perturbative expansion forms a geometric series to all orders, has
convergent behavior.
\end{enumerate}

Virtually any perturbative QCD prediction can be written in the form of
a commensurate scale relation, thus eliminating any uncertainty due to
renormalization scheme or scale dependence.  Recently it has been
shown\cite{BPT} how the commensurate scale relation between the
radiative corrections to $\tau$-lepton decay and $R_{e^+e^-}(s)$ can be
generalized and empirically tested for arbitrary $\tau$ mass and nearly
arbitrarily functional dependence of the $\tau$ weak decay matrix
element.

An essential feature of the \av(Q) scheme is the absence of any
renormalization scale ambiguity, since $Q^2$ is, by definition, the
square of the physical momentum transfer.  The \av\ scheme naturally
takes into account quark mass thresholds, which is of particular
phenomenological importance to QCD applications in the mass region close
to threshold.  As we have seen, commensurate scale relations provide an
analytic extension of the conventional \ms\ scheme in which many of the
advantages of the \av\ scheme are inherited by the \amst\ scheme, but
only minimal changes have to be made.  Given the commensurate scale
relation connecting \amst\ to \av\, expansions in \amst\ are effectively
expansions in \av\ to the given order in perturbation theory at a
corresponding commensurate scale.

The calculation of $\psi_V^{(1)}$, the two-loop term in the Gell-Mann
Low function for the $\alpha_V$ scheme, with massive quarks gives for
the first time a gauge invariant and renormalization scheme independent
two-loop result for the effects of quark masses in the running of the
coupling.  Renormalization scheme independence is achieved by using the
pole mass definition for the ``light" quarks which contribute to the
scale dependence of the static heavy quark potential.  Thus the pole
mass and the $V$-scheme are closely connected and have to be used in
conjunction to give reasonable results.

It is interesting that the effective number of flavors in the two-loop
coefficient of the Gell-Mann Low function in the $\alpha_V$ scheme,
$N_{F,V}^{(1)}$, becomes negative for intermediate values of $Q/m$.
This feature can be understood as anti-screening from the non-Abelian
contributions and should be contrasted with the QED case where the
effective number of flavors becomes larger than one for intermediate
$Q/m$.  For small $Q/m$ the heavy quarks decouple explicitly as expected
in a physical scheme, and for large $Q/m$ the massless result is
retained.

The analyticity of the $\alpha_V$ coupling can be utilized to obtain
predictions for any perturbatively calculable observables including the
finite quark mass effects associated with the running of the coupling.
By employing the commensurate scale relation method, observables which
have been calculated in the $\overline{\mbox{MS}}$ scheme can be related
to the analytic V-scheme without any scale ambiguity.  The commensurate
scale relations provides the relation between the physical scales of two
effective charges where they pass through a common flavor threshold.  We
also note the utility of the \av\ effective charge in supersymmetric and
grand unified theories, particularly since the unification of couplings
and masses would be expected to occur in terms of physical quantities
rather than parameters defined by theoretical convention.

As an example, the finite quark mass corrections connected with the
running of the coupling for the non-singlet hadronic width of the
Z-boson have been calculated in the analytic V-scheme and compared with
the standard treatment in the $\overline{\mbox{MS}}$ scheme.  The
analytic treatment in the V-scheme gives a simple and straightforward
way of incorporating these effects for any observable.  This should be
contrasted with the $\overline{\mbox{MS}}$ scheme where higher twist
corrections due to finite quark mass threshold effects have to be
calculated separately for each observable.  The V-scheme is especially
suitable for problems where the quark masses are important such as for
threshold production of heavy quarks and the hadronic width of the
$\tau$ lepton.

It has now also been shown that the NLO corrections to the highest
eigenvalue of the BFKL equation become controllable and meaningful
provided one uses physical renormalization scales and schemes relevant
to non-Abelian gauge theory.  BLM optimal scale setting automatically
sets the appropriate physical renormalization scale by absorbing the
non-conformal $\beta$-dependent coefficients.  A striking feature of the
NLO BFKL Pomeron intercept in the BLM/CSR approach is its very weak
$Q^2$-dependence, which provides approximate conformal invariance.
These new results open new windows for applications of NLO BFKL
resummation to high-energy phenomenology, particularly virtual
photon-photon scattering.

\section*{Acknowledgments}
Much of this work is based on collaborations
with Michael Melles, Victor Fadin, Mandeep Gill, Lev Lipatov, Andrei
Kataev, Victor Kim, Gregory Gabadadze, and Grigorii B. Pivovarov.  We
also thank Yitzhak Frishman, Georges Grunberg, Einan Gardi, and Marek
Karliner for helpful conversations.  SJB also thanks the organizers of
the La Thuile conference, particularly Professor Mario Greco, for their
outstanding hospitality.

\end{document}